# Dual Action of Electrolyte Additives in High-Voltage Li-Ion Cells

Aidin I. O'Brien, Bingning Wang, Stephen E. Trask, Adam Tornheim, Chen Liao, Marco-Tulio F. Rodrigues

Chemical Sciences and Engineering Division, Argonne National Laboratory, Lemont, IL, USA

**Contact:** marco@anl.gov

**Abstract**

Electrolyte additives have outsized effects on cell performance, but determining their mechanism of action can be difficult. Here, we evaluate whether simple electrochemical tests can reveal which electrode (cathode, anode or both) is targeted by specific additives. Our experiments quantified the rates of reductive and oxidative side-reactions in NMC532 vs. graphite cells containing 12 different electrolyte formulations. We found that gains in interphasial stability were highly correlated, with successful additives decreasing both types of side-reactions by similar magnitude. This "dual action" could be indicative of the role of crosstalk on cell degradation, as soluble species formed on one electrode can decompose at the other end of the cell. Although electrochemical tests are insufficient for understanding mechanisms of improvement, they can reveal interesting effects that complement other characterization tools.

## Introduction

The performance of Li-ion batteries is strongly influenced by phenomena at the interface between the electrodes and the electrolyte.[1] This has important design implications, as small amounts of chemicals displaying preferential electrochemical activity can have outsized effects on cell behavior. Accordingly, electrolyte additives are perhaps the most cost-effective way of imparting change to the cell. The primary goal of many additives is to cast protective layers on the electrodes to prevent parasitic processes. Much of the capacity fade in Li-ion cells arises from reduction side-reactions at the surface of the anode, which can form or reconstitute the solid electrolyte interphase (SEI). The technological and commercial success of Li-ion batteries can largely be attributed to the development of electrolyte components that can form a better SEI and thus extend cell life.[2] Electrolyte species can also affect the cathode, as unwanted oxidation can lead to impedance rise and facilitate the leaching of transition metal ions.[3-5]

While the performance of new electrolyte formulations can be readily verified, establishing their exact mechanism of action is challenging. Chemical intuition has been a helpful tool for identifying candidate additives, indicating that, to some extent, an understanding between target activity and molecular features already exists. Many additives known to be effective "SEI-formers", such as VC, PES and tVCBO (see Figure 1 for structures and Table 1 for full chemical names), contain unsaturated bonds that can be reduced to initiate polymerization. These structural features also decrease the energy of their lowest unoccupied molecular orbital, so that they react preferentially over carbonate solvents and establish compositional dominance at the SEI despite being present in small quantities.[6] Other additives, like PyDMA and LiDFP, possess high-energy occupied orbitals arising from nitrogen lone pairs and anionic oxygen lone pairs, respectively, that make them prone to electrochemical oxidation at the cathode to yield protective layers.[7, 8] There are also molecules like TMSPi and bis(trimethylsilyl) malonate, which are transformed within the electrolyte through spontaneous reactions with $LiPF_6$, generating *in situ* the active molecules that improve cell performance.[9] Additionally, many structural moieties like the silyl groups present in MS, TMSPi and TMSML, are expected to react with rogue fluoride in the electrolyte, trapping HF that could attack the cathode surface and, through crosstalk, accelerate side-reactions at the anode.[10, 11] Substantial literature exists linking the activity of these electrolyte components to either the cathode or anode, or both (Table 1). This assignment is typically done by correlating evidence collated from various characterization methods with cycling performance.

The present work was motivated by an interest in assigning the electrode-specific activity of electrolyte formulations solely using electrochemical testing. Our approach is based on the fact that parasitic reduction and oxidation are independently quantifiable and occur only at the anode and cathode, respectively.[12, 13] Thus, favorable interactions between electrolyte components and a given electrode should specifically mitigate the associated type of side-reaction. We analyzed 12 different electrolyte formulations tested in $LiNi_{0.5}Mn_{0.3}Co_{0.2}O_2$ (NMC532) vs. graphite cells exposed to high voltages (4.4 V). Our baseline ("Gen2") formulation consisted of 1.2M $LiPF_6$ in 3:7 wt:wt EC:EMC (that is, ethylene carbonate and ethyl methyl carbonate). We also tested a composition with decreased EC content ("low EC", where solvent ratios were 1:9 by volume) as ethylene carbonate is prone to oxidation side-reactions.[3, 4] The other ten electrolyte formulations (Figure 1) involved the use of various salts and molecules as additives to Gen2, at concentrations specified in Table 1. We report that improvements in oxidative and reductive stability were highly correlated, suggesting that electrolyte activity at the interphases of both electrodes may be intrinsically connected.

Table 1. Name and concentration of additives used in this study, and their expected activity based on prior literature. Assignment of improved interphases at the cathode, anode or both are not mutually exclusive, and just reflect the scope of the work reported in each reference. For MS and TMSPi, the duration for which the *in situ* reactions were allowed to proceed before cell assembly is also included.

| Additive (abbreviation) | Additive (full name) | Concentration | Anode activity | Cathode activity | Dual activity |
|---|---|---|---|---|---|
| FEC | fluoroethylene carbonate | 3 wt% | ref. [14] | | ref. [15, 16] |
| LiDFP | lithium difluorophosphate | 0.1 M | | ref. [8, 17] | ref. [18, 19] |
| LiDFOB | lithium difluoro(oxalato)borate | 1 wt% | | | ref. [20, 21] |
| MS | bis(trimethylsilyl) malonate | 1 wt% (2 days) | | ref. [22, 23] | |
| PES | prop-1-ene-1,3-sultone | 1 wt% | | | ref. [24-26] |
| PyDMA | 1-(dimethylamino) pyrrole | 0.5 wt% | | ref. [7] | |
| TMSML | 4-(trimethylsilyl) morpholine | 1 vol% | | ref. [11] | |
| TMSPi | tris(trimethylsilyl) phosphite | 2 wt% (1 week) | ref. [27] | ref. [9, 10, 28] | |
| tVCBO | trivinylcyclotriboroxane | 0.25 wt% | ref. [29] | ref. [30] | |
| VC | vinylene carbonate | 2 wt% | ref. [14] | | ref. [15, 31-33] |

***Experimental***

***Electrode composition and cell assembly.*** All electrodes used in this study were fabricated at Argonne's Cell Analysis, Modeling and Prototyping (CAMP) Facility using a roll-to-roll coater. The cathode comprised 90 wt% NMC532 (Toda), 5 wt% C45 conductive carbon (Timcal) and 5 wt% polyvinylidene fluoride binder (PVDF, Solvay 5130) coated on 20 μm battery grade aluminum foil. The 9.13 mg/cm$^2$ coating was calendered to 34 μm to achieve 33.9% porosity. The anode consisted of 91.83 wt% graphite (SLC1506T, Superior Graphite), 2 wt% C45, 6 wt% PVDF (9300, Kureha) and 0.17 wt% oxalic acid, coated on 10 μm battery grade copper foil. The 6.37 mg/cm$^2$ coating was calendered to 47 μm (37.5% porosity). Tests were performed in 2032-format coin cells containing 40 μL of electrolyte, and using 14, 15 and 16 mm discs for the cathode, anode and Celgard 2500 separator, respectively. Cells were assembled in an argon-filled glovebox.

***Electrolytes.*** The baseline "Gen2" electrolyte was 1.2M $LiPF_6$ dissolved in 3:7 wt:wt EC:EMC, and was supplied by Tomiyama. Most electrolytes were prepared by adding appropriate amounts of the electrolyte additive to the baseline (Table 1), with specific quantities established based on prior work at Argonne. A variant of the baseline with lower EC content was also prepared and consisted of 1.2M $LiPF_6$ dissolved in 1:9 v:v EC:EMC. For the latter, individual components were sourced from Novolyte Technology. Additives were purchased from Solvay (FEC), Hunan Chemfish Pharmaceutical Co., Ltd. (LiDFP), TCI America (PES and VC), ACS Scientific, Inc. (tVCBO) and Sigma Aldrich (PyDMA, TMSPi and TMSML), and were used as received. LiDFOB was produced at Argonne's Materials Engineering Research Facility (MERF).[34] MS was formed *in situ* in the electrolyte upon spontaneous reaction between bis(trimethylsilyl) malonate (TCI America) and $LiPF_6$ for two days.[23] TMSPi (Sigma Aldrich) was added to the baseline electrolyte and used after aging for 1 week,[9] as it reacts with $LiPF_6$ to form fluorinated compounds reported to be beneficial to the cell.

***Test protocol.*** Cells were placed in a 30 °C environmental chamber and allowed to rest for 2 hours for thermal equilibration. A tap charge to 1.5 V was applied for 15 minutes, followed by 6 hours of rest. The initial three cycles were at 230 μA (~C/10), followed by a voltage hold at 4.05 V for 15 hours (Figure 2). This hold was implemented to assist with saturation of the anode overhang early in the experiment, decreasing its apparent contributions to the quantified parasitic reduction.

At 4.05 V, the anode is at the beginning of the graphite plateau at ~89 mV vs. $Li/Li^+$, while the cathode is maintained at relatively low potentials to prevent premature surface reactivity. Apart from this conditioning step, all cycling is performed between 2.9 and 4.4 V.

The main testing backbone consisted of 30 constant current cycles at 147 μA (~C/17), which is slightly below the 150 μA maximum for the lowest current range of the Maccor 4000 cycler used in these tests.[13] Maintaining the cycler at that lower range is needed to minimize sensing errors when computing the rates of parasitic reactions, as we discuss later. After these 30 cycles, cells were charged to 4.4 V at 147 μA and held at that voltage for 100 hours to quantify parasitic oxidation (see below). Throughout the test, all charge half-cycles were followed by 335 seconds of rest, with data for the initial periods of voltage relaxation being recorded every 0.1 s; the resulting data was used as a proxy for cell impedance.

***Conventions.*** Cell formation was assumed to encompass the three ~C/10 cycles, the voltage hold at 4.05 V and the initial two cycles at 147 μA. The seventh cycle was taken as the "beginning of life" (BOL, Figure 2) and was used to define properties of interest (such as initial capacity and capacity loss). Most data discussed in the present work will focus on the 28 cycles between the BOL and the 147 μA charge leading to the high-voltage hold. When relevant, figures will present the average and standard deviation for triplicate cells, and many quantities are normalized per gram of NMC532.

***Impedance.*** Cell impedance was assumed to be proportional to the decrease in open circuit voltage (OCV) measured 108 milliseconds after the end of charge. This semi-quantitative correlation is possible due to the flat voltage profile of graphite, which forces the cathode to experience a constant end-of-charge potential throughout much of the cell's life. An increase in polarization in either electrode will prompt premature termination of the charging half-cycle, which can be diagnosed by the cell relaxing to a lower OCV due to less delithiation of the cathode.

***Quantification of parasitic reactions.*** Three methods were used to quantify the extent of side-reactions: endpoint slippage, $CE/CR$ and voltage hold. In all cases it was assumed that loss of active material was negligible during aging, which is reasonable for the electrodes and conditions used in this study.[35]

Endpoint slippage is based on tracking the shifts of the termination of charge and discharge half-cycles (i.e., the endpoints) when testing data is represented along a cumulative capacity axis.[12] In graphite cells, slippage of discharge and charge endpoints can be uniquely correlated with the quantity of parasitic reduction and oxidation, respectively, that the cell has experienced.[12] For ease of computation, the slippage was calculated per cycle and then summed to indicate the cumulative capacities consumed by each type of side-reaction. As defined in ref. [12], the discharge endpoint slippage accrued within a cycle $k$ is $D_{slip,k} = Q_{c,k} - Q_{d,k}$, where $Q_c$ and $Q_d$ are the measured charge and discharge capacities. The charge endpoint slippage from that same cycle is given by $C_{slip,k} = Q_{c,k} - Q_{d,k-1}$, where $Q_d$ is taken at the preceding cycle $k-1$. Cumulative slippages are obtained from summation of their cycle-by-cycle counterparts.

The $CE/CR$ method was developed by Tornheim and O'Hanlon (ref. [13]) and quantifies side-reactions in terms of parasitic currents per cycle $I_{red,k}$ and $I_{ox,k}$ for reduction and oxidation, respectively. Here, $I_{red,k} = I_k(1 - CE_k)/(1 + CE_k)$ and $I_{ox,k} \approx I_k(CR_k - CE_k)/(CR_k + CE_k)$, where $CE_k$ and $CR_k$ are the measured coulombic efficiency at cycle $k$ and capacity retention with respect to cycle $k-1$ when the cell is tested under the constant current $I_k$. For comparison with the endpoint slippage method, the parasitic currents were converted to capacities through multiplication by the duration $\tau_k$ of each full cycle.

The voltage hold technique assumes that, for specific cell configurations, the current measured after relaxation is proportional to the instantaneous rate of either parasitic oxidation or reduction. The side-reaction quantified by this method takes place at a working electrode that is paired against a counter-electrode presenting a flat voltage profile.[3, 36] Under these conditions, keeping the cell voltage constant will also maintain the working electrode at a constant state of lithiation, minimizing contributions from tapering to the measured current. The requirement of a constant potential at the counter-electrode makes configurations vs. graphite or LTO ideal for measuring parasitic oxidation at the cathode,[3] and vs. LFP for assessing parasitic reduction at the anode.[36] Hence, in the NMC532 vs. graphite cells analyzed here, the voltage hold quantifies parasitic oxidation as cells are maintained at 4.4 V.

## *Results and Discussion*

***Cycling.*** The discharge capacities and cumulative capacity losses measured during testing are shown in Figure 3. Only data from cycle 7 onwards is presented, as the preceding cycles were used for conditioning of the cells (Figure 2). The shading indicates the standard deviation for triplicate cells tested for each electrolyte system, and the color code is maintained throughout the entire manuscript for ease of visualization.

This dataset shows that nearly all electrolyte formulations decreased the initial discharge capacity of the cells relative to the Gen2 baseline (Figure 3a, shown in gray), and led to limited mitigation of capacity loss (Figure 3b). This is shown more clearly in Figure 4a, where the gray bands indicate the standard deviations for Gen2. The overall amount of capacity fade appeared to be correlated with the coulombic efficiency exhibited by the cells at the end of the test (Figure 4b), suggesting that loss of $Li^+$ inventory is a dominant source of degradation. TMSML is a clear exception to these trends, as it experienced little capacity loss while displaying low coulombic efficiency. This system exhibited signs of experiencing extensive side-reactions within conditions explored in this work, with strong indication of the presence of a redox shuttle; evidence will be examined below. This system will be an outlier for many of the tests.

At a surface level, some of the trends in Figure 3 seem to indicate that electrolyte additives offered little value to cell performance. As we discuss later, this early impression can be dispelled upon further inspection of the data.

***Formation.*** Figure 3 and Figure 4 indicated that most systems presented decreased initial capacities relative to the baseline; we now investigate the origin of this gap. Figure 5 shows dQ/dV traces within the voltage range where early redox processes are visible during the first cell charge. Given the low cathode potentials at this interval, these events are likely indicative of reactions at the anode surface to form the SEI. The baseline presented a broad feature with onset at ~2.4 V and a peak centered at ~3 V (Figure 5a). These have commonly been attributed to reduction of ethylene carbonate.[34, 37, 38] Some electrolyte systems (low EC, LiDFP and tVCBO) presented little changes relative to the baseline, possibly indicating a limited role in the early SEI formation; this is notable, as the two latter additives have been reported to improve the SEI (Table 1). Other cases, like LiDFOB, MS, PyDMA and TMSML, exhibited additional redox events at low voltages while still maintaining features associated with EC decomposition. In these cases, the additive had direct contributions to the SEI but that were nonetheless insufficient to completely suppress additional reactivity. With TMSPi, an increased amplitude is observed for the peak at ~3 V, suggesting that the additive might also decompose at that voltage. Finally, FEC, PES and VC appeared to be involved in reduction reactions that prevented significant degradation of ethylene carbonate, in agreement with prior reports.[14, 39, 40]

The lower BOL capacity for many systems cannot be explained simply by the extent of the first-cycle reactivity to form the SEI: addition of LiDFP to the electrolytes leads to decreased capacities (Figure 4a) while exhibiting dQ/dV traces that are identical to the baseline (Figure 5e), while the higher initial reactivity of PyDMA (Figure 5h) did not result in a lower initial capacity (Figure 4a). We then considered that inefficiencies may accumulate during the six initial conditioning cycles that are not fully captured by the redox processes related to SEI formation in Figure 5. Figure 6a presents the cumulative capacity lost in these cycles, calculated by subtracting the discharge capacity from the charge capacity for each cycle and summing the six differences; it also shows the beginning-of-life (BOL) discharge capacity measured in cycle 7. No clear trend is observed, suggesting that SEI growth is not the only source of decreased initial capacity in the various systems.

We also considered that certain additives could affect cell capacity by increasing cell impedance through high-resistance layers formed on the cathode and/or anode. In Figure 6b, we use the voltage drop measured 108 milliseconds after the end of charging as a proxy for resistance. This is also unsuccessful in explaining the observations, as systems such as LiDFOB, MS and PES present improved initial resistances without benefitting from higher capacities. The factors determining how the additives affect the initial cell capacity are likely complex, involving the features mentioned above and, perhaps, additional ones we have not considered.

***Quantifying parasitic reactions.*** Revisiting Figure 3b and Figure 4, no additives presented substantial mitigation of capacity loss relative to the baseline electrolyte. (With the clear exclusion of the suspicious TMSML; more on that below.) We now show that several of the electrolyte formulations we investigated are actually highly effective.

In much of the prior literature, "loss of $Li^+$ inventory" (LLI) is conventionally understood as the decrease in cell capacity due to parasitic reactions. As discussed elsewhere,[12] LLI is more rigorously defined as the net balance between irreversible reduction and oxidation reactions. The former is associated with processes at the anode that consume electrons from the cell, like SEI growth. The latter involves events leaving a surplus of electrons at the cathode, like EC oxidation and lattice oxygen loss.[3] Under this perspective, capacity loss depends on the balance between how much $Li^+$ is gained (from oxidation side-reactions) and how much is lost (from reduction).

These side-reactions can be quantified electrochemically by tracking the slippage of endpoints.[12] This method involves the analysis of the termination of charge and discharge half-cycles when testing data is represented across a cumulative capacity axis. Figure 7a,b shows the slippages of the discharge and charge endpoints, respectively, accumulated throughout the test; the final slippages are summarized in Figure 8a. A positive effect from some of the additives is evident, with FEC, LiDFP, MS, PES, TMSPi and VC showing decreased levels of both parasitic oxidation and reduction relative to the baseline. Below, we examine three points of interest relating to the values in Figure 8a.

The first is the apparent contradiction between the stability suggested by Figure 8a and the persistent capacity losses in Figure 4a. The reason becomes clear once one notes that the experimental points in Figure 8a are distributed along a line that runs parallel to the main diagonal, suggesting that decreasing the slippage of one of the endpoints leads to a change of similar magnitude on the other. That is, an electrolyte that prevents the loss of X mAh/g due to a better SEI will also forgo a gain of ~X mAh/g from oxidation due to less reactivity at the cathode, resulting in net capacity fade that does not differ much from that of the baseline. This is directly seen in Figure 8b, which presents in its vertical axis the difference between the cumulative discharge and charge slippages for all samples, with most formulations neighboring the baseline. This quantity equals the net capacity loss experienced by the cell, as verified by comparison with Figure 4.

Perhaps more interesting is the representation of the summed slippages in the horizontal axis of Figure 8b, indicating the *overall* reactivity of each electrolyte. Parasitic oxidation can also involve decomposition of electrolyte species, providing a temporary capacity gain at the expense of irreversible electrolyte loss and long-term cell degradation.[41] Desirable electrolytes would ideally fall within the lower left area of Figure 8b, combining decreased reactivity with lower capacity fade. Unfortunately, none of the systems studied here are situated on that quadrant. There are, however, several additives that can decrease the total electrolyte reactivity without major compromises in capacity retention, such as MS, PES, TMSPi and VC. Even without gains in capacity retention, cells containing these additives are expected to display better longevity than the baseline when tested under lean electrolyte conditions. Data visualization schemes like Figure 8b, juxtaposing capacity loss ($D_{slip} - C_{slip}$) and total parasitic capacity ($D_{slip} + C_{slip}$), offer a convenient way of judging the effectivity of electrolytes.

The second point of interest in Figure 8a is why the data points follow a line parallel to the main diagonal. There is evidence suggesting that ineffective SEIs favor the sustained production of soluble reduction products, which can react at the cathode surface and lead to impedance rise.[42] With better anode passivation, formation of these reactive species is frustrated and improved performance is obtained. This "crosstalk" could explain the linked reactivity observed for both electrodes: mitigating reduction reactions will also prevent certain parasitic oxidation processes from happening. It is interesting to note that, except for LiDFP, additives capable of mitigating side-reactions participated at the early stages of SEI formation (compare Figure 8a and Figure 5). It is also conceivable that the opposite route of stabilization is also true, where a more protective surface layer on the cathode could minimize the amount of Mn dissolution, thus preventing the acceleration of SEI growth on graphite.[5, 10]

A final point to be highlighted in Figure 8a is the sheer level of reactivity exhibited by TMSML. The elevated and similar magnitudes for charge and discharge endpoint slippage suggest that a redox shuttle is present. A redox shuttle involves molecules that will engage in reversible cycles of oxidation and reduction, causing high levels of specific reactivity with no net changes to the $Li^+$ inventory.[43] Additionally, TMSML is also to the left of the trendline in Figure 8a, showing that it experienced a higher increase in parasitic oxidation than in reduction with respect to the baseline. The resulting excess $Li^+$ inventory from persistent side-reactions explains why capacities measured for this system remained nearly unchanged during testing (Figure 3 and Figure 4).

***Comparative quantification of side-reaction rates.*** Figure 7 illustrates the use of endpoint slippages to quantify the extent of side reactions. Here we compare those findings with results obtained from other methods that can be applied to our data.

The first one is the $CE/CR$ method developed by Tornheim and O'Hanlon,[13] which uses coulombic efficiency and the cycle-by-cycle capacity retention to calculate average parasitic currents $I_{red}$ and $I_{ox}$ for reduction and oxidation, respectively. In a previous work,[12] it has been shown that this method is mathematically equivalent to the analysis of endpoint slippages and should thus result in equal reactivity. This is verified in Figure 9, where the endpoint slippages measured within a single cycle are compared to the parasitic currents multiplied by the total time $\tau$ elapsed in a cycle to convert $I_{red}$ and $I_{ox}$ into capacities. Figure 9 shows such data for all cycles of all systems, and every single one falls perfectly along the main diagonal, proving that the equivalence expected for both methods is indeed observed experimentally. Our analysis also showed that both methods require data manipulation that produces similar levels of error propagation (not shown here for brevity), suggesting that the choice between them is largely a matter of convenience.

Parasitic reactions can also be quantified using voltage holds as the cells are held at 4.4 V for 100 hours after cycling (Figure 2). During the voltage hold, the current decayed sharply during the initial 10 hours (Figure 10a) as NMC/graphite cells relax notably fast.[36] The integrated capacities exchanged during the hold (Figure 10b) presented a smoother profile, facilitating quantitative analysis. The rates of parasitic oxidation were determined from the slope of a linear fit to the final 30 hours of data (Figure 10b). Figure 10c compares the resulting values with the calculated charge endpoint slippages, showing a good qualitative agreement between the methods for most samples. A few deviations are present, as in the case of PES that shows charge slippage identical to TMSPi and VC but a visibly higher hold slope. The lack of perfect agreement between the two techniques is unsurprising, as they rely on different stimuli to track parasitic oxidation (a full cycle vs. a potentiostatic hold) and report distinct quantities (cumulative vs. instantaneous oxidation, respectively). Extended exposure to high electrode potentials during the voltage hold may penalize additives displaying certain types of reactivity at the cathode surface. Importantly, all methods correctly identify TMSML as an outlier.

***Mechanism of function of electrolyte systems.*** In Table 1 we attempted to summarize the expected target activity for the different electrolyte systems based on prior literature. Since assigning beneficial effects to a single electrode is difficult, "activity" was considered based on claims made in the original publications and/or on the experimental observation of surface films with composition that can be traced back to the additive. These assignments can be compared to the changes in reactivity we report in Figure 8. Many of the entries in Table 1, like FEC, LiDFP and PES, are expected to benefit both electrodes. This broadly agrees with Figure 8, where such systems tended to display lowering in both charge and discharge slippage. Notable exceptions are LiDFOB and PyDMA, for which the improvements reported in the literature in other systems were not reproduced despite their obvious role in initial SEI formation (Figure 5). On the other hand, MS displayed a decrease in discharge endpoint in addition to the expected activity in the cathode, suggesting an expanded role of the additive. The "low EC" composition was also an interesting case, as the SEI stability deteriorated without the gains in oxidative stability reported in the literature.

Revisiting the arrangement of the data points in Figure 8a along the diagonal line, it seems to be difficult to ascribe causation of benefits from electrochemical data alone. As we have shown above, additives that outperformed the baseline (in terms of minimizing total parasitic capacities) tended to do so by simultaneously mitigating side-reactions at both the cathode and anode. At face value, that could indicate that successful additives would have a direct protective role on both electrodes. This, however, may not necessarily be true, as mitigating side-reactions in one electrode may prevent the generation of reactive species that can travel through the cell and react at the other electrode. Given this symmetry of outcomes at both electrodes, it is difficult to derive mechanistic insight from electrochemical data alone.

***A note on experimental error.*** Much of the discussion in the present work is based on the experimental quantification of small signals. The important work by Smith et al. (ref. [44]) has shown that dedicated hardware may be required for measuring capacity and coulombic efficiency within ~0.01%. Our tests were not performed using a high-precision cycler, resulting in a theoretical uncertainty of ~0.1% for capacity and coulombic efficiency (see error propagation analysis in the Appendix). Nevertheless, this value likely overestimates the effects of sensing accuracy on the data, as it assumes that maximum distortions in current and voltage will always coincide. A concrete example of that is the error propagated into the discharge endpoint slippage measured within a single cycle. The estimate provided in the Appendix (3 μAh, or ~0.2 mAh/g) is significantly larger than the standard deviation observed for all our tests (Figure A1). While we recommend that precision instrumentation be used when accessible, the internal consistency of the data we collected across several cells and channels appears to suggest that the specifications for our testing hardware could have been conservative. It is also possible that errors are still significant but systematic and thus have a limited impact on our comparative analysis of electrolyte reactivity.

***Conclusions.*** Twelve different electrolyte formulations were tested in Li-ion cells exposed to high voltages. We found that electrolyte additives generally failed to improve the initial capacity or the capacity retention measured with the Gen2 baseline. Nevertheless, this ineffectiveness was just apparent, as further analyses indicated that certain additives successfully decreased the rates of side-reactions.

We compared three electrochemical methods for quantifying parasitic processes: endpoint slippage, CE/CR and voltage hold. The first two approaches provide the rates of reduction and oxidation experienced during a full cycle, while the latter probe the instantaneous parasitic oxidation currents at a specific cell voltage. Endpoint slippage and CE/CR had been previously demonstrated to be mathematically equivalent, for which we have now presented experimental proof. Trends obtained using voltage holds were similar to the measured by the other techniques but not identical, as the former is especially sensitive to systems displaying more reactivity at high voltages.

The central motivation of the present work was to evaluate whether simple electrochemical tests could assist in identifying which electrode is being targeted by electrolyte additives. We found that changes in electrolyte formulation led to a similar decrease in the rate of both oxidation and reduction side-reactions. We hypothesized that this concerted improvement is due to mitigation of crosstalk, as better passivation of one interphase prevents the formation of reactive soluble components that can react on the other electrode. Overall, electrochemical tools can only quantify effects rather than causes, making them partially ineffective in providing direct mechanistic understanding.

It should be emphasized that the electrolyte performance discussed here must be considered within the context of the electrodes and testing conditions from our experiments. Nevertheless, we believe that our approach and reasoning can be useful to other researchers interested in acquiring a deeper understanding of the activity of electrolyte additives.

***Acknowledgements***

Support from the Transportation Technologies Office (TTO) of the U.S. Department of Energy (DOE) is gratefully acknowledged. This work was supported in part by the U.S. Department of Energy, Office of Science, Office of Workforce Development for Teachers and Scientists (WDTS) under the Science Undergraduate Laboratory Internships Program (SULI). The submitted manuscript has been created by UChicago Argonne, LLC, Operator of Argonne National Laboratory ("Argonne"). Argonne, a U.S. Department of Energy Office of Science laboratory, is operated under Contract No. DE-AC02-06CH11357. The U.S. Government retains for itself, and others acting on its behalf, a paid-up nonexclusive, irrevocable worldwide license in said article to reproduce, prepare derivative works, distribute copies to the public, and perform publicly and display publicly, by or on behalf of the Government.

***Appendix***

Consider a 2.45 mAh cell cycling at C/17. For a Maccor 4000 at its lowest current range, the accuracy for current and voltage are $\delta_I = 30\ nA$ and $\delta_V = 2\ mV$. Effects on capacity from current fluctuation are maximum when the current is higher or lower by $\delta_I$ for the entire duration of a half-cycle, and this upper bound is expressed by

$$\delta Q_I = \delta_I \cdot 17h = 0.51\ \mu Ah$$

For the voltage, early or late termination of charge $(c)$ and discharge $(d)$ by $\delta_V$ would cause a change in cell capacity that is proportional to the slope of the voltage profile and the end of charge (EOC) or discharge (EOD):

$$\delta Q_{V,c} = \delta_V \cdot \left.\frac{dQ}{dV}\right|_{EOC} = \delta_V \cdot 0.00144\ {}^{Ah}\!/_{V} = 2.9\ \mu Ah$$

$$\delta Q_{V,d} = \delta_V \cdot \left|\left.\frac{dQ}{dV}\right|_{EOD}\right| = \delta_V \cdot 0.00015\ {}^{Ah}\!/_{V} = 0.3\ \mu Ah$$

For our NMC532 vs. graphite cells, the uncertainty in the capacity measurements is higher during charge half-cycles by an order of magnitude as voltage profiles are much steeper around the EOD (i.e., the $dQ/dV$ is smaller at that point). The $\delta Q_{V,c}$ is also significantly larger than $\delta Q_I$.

Assuming that errors from current and voltage are independent, their combined effect on cell capacity is

$$\delta Q_c = \sqrt{\delta {Q_I}^2 + \delta {Q_{V,c}}^2} = 2.9\ \mu Ah$$

$$\delta Q_d = \sqrt{\delta {Q_I}^2 + \delta {Q_{V,d}}^2} = 0.6\ \mu Ah$$

resulting in relative errors of 0.12% and 0.02% for charge and discharge, respectively.

For reference, we can also assume that $\delta Q_c$ and $\delta Q_d$ are independent, resulting in a relative error for measuring the coulombic efficiency CE equal to

$$\frac{\delta \mathrm{CE}}{CE} = \sqrt{\left(\frac{\delta Q_d}{Q_d}\right)^2 + \left(\frac{\delta Q_c}{Q_c}\right)^2} \approx 0.001$$

In other words, CE measurements are, at worst, off by $\pm 0.1\%$.

Considering the error propagation for the discharge endpoint slippage $D_{slip}$ within a single cycle, we get

$$\delta D_{slip} = \sqrt{\delta {Q_c}^2 + \delta {Q_d}^2} = 3\ \mu Ah$$

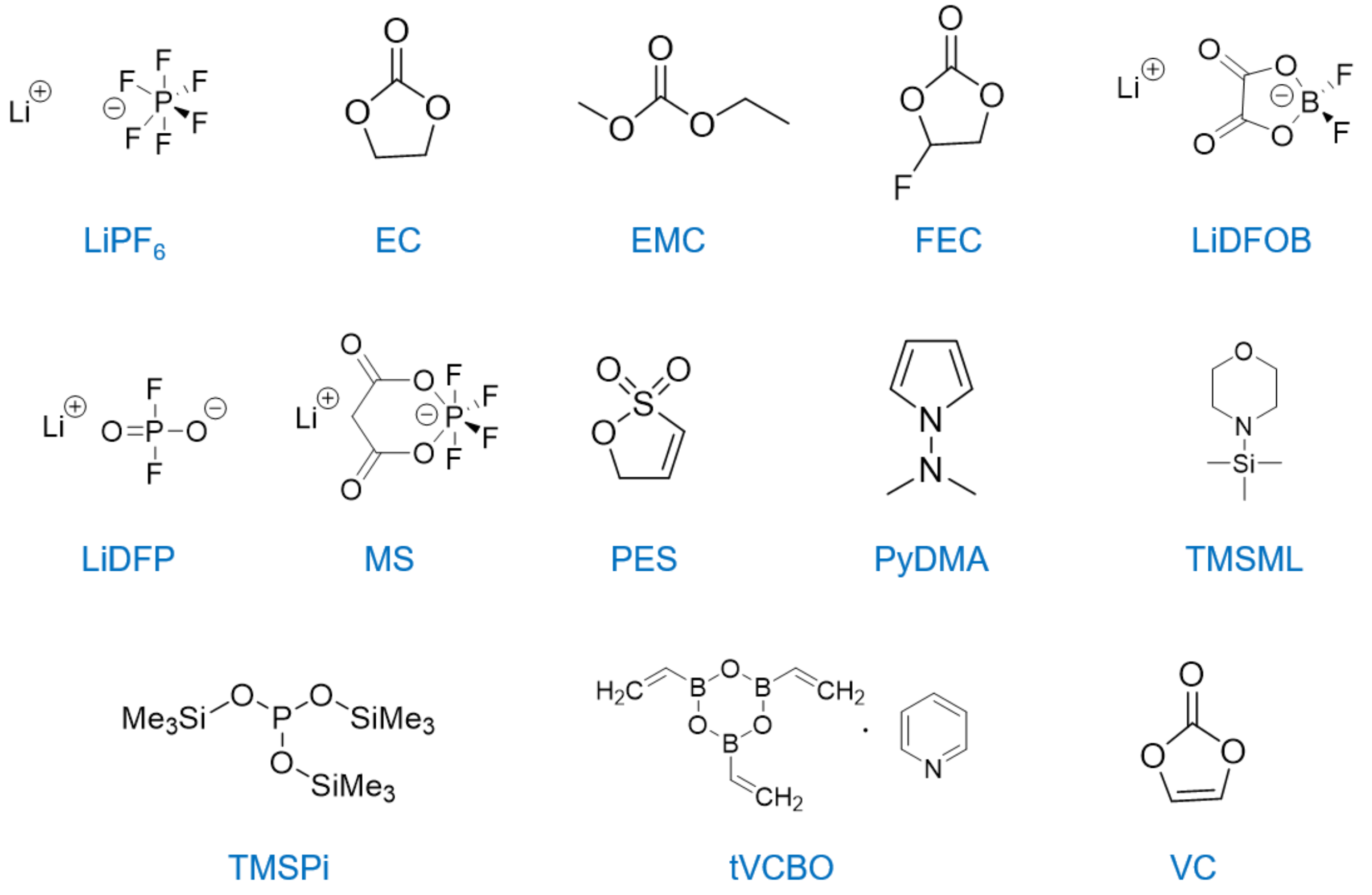

Figure 1. Electrolyte components used in the evaluated formulations.

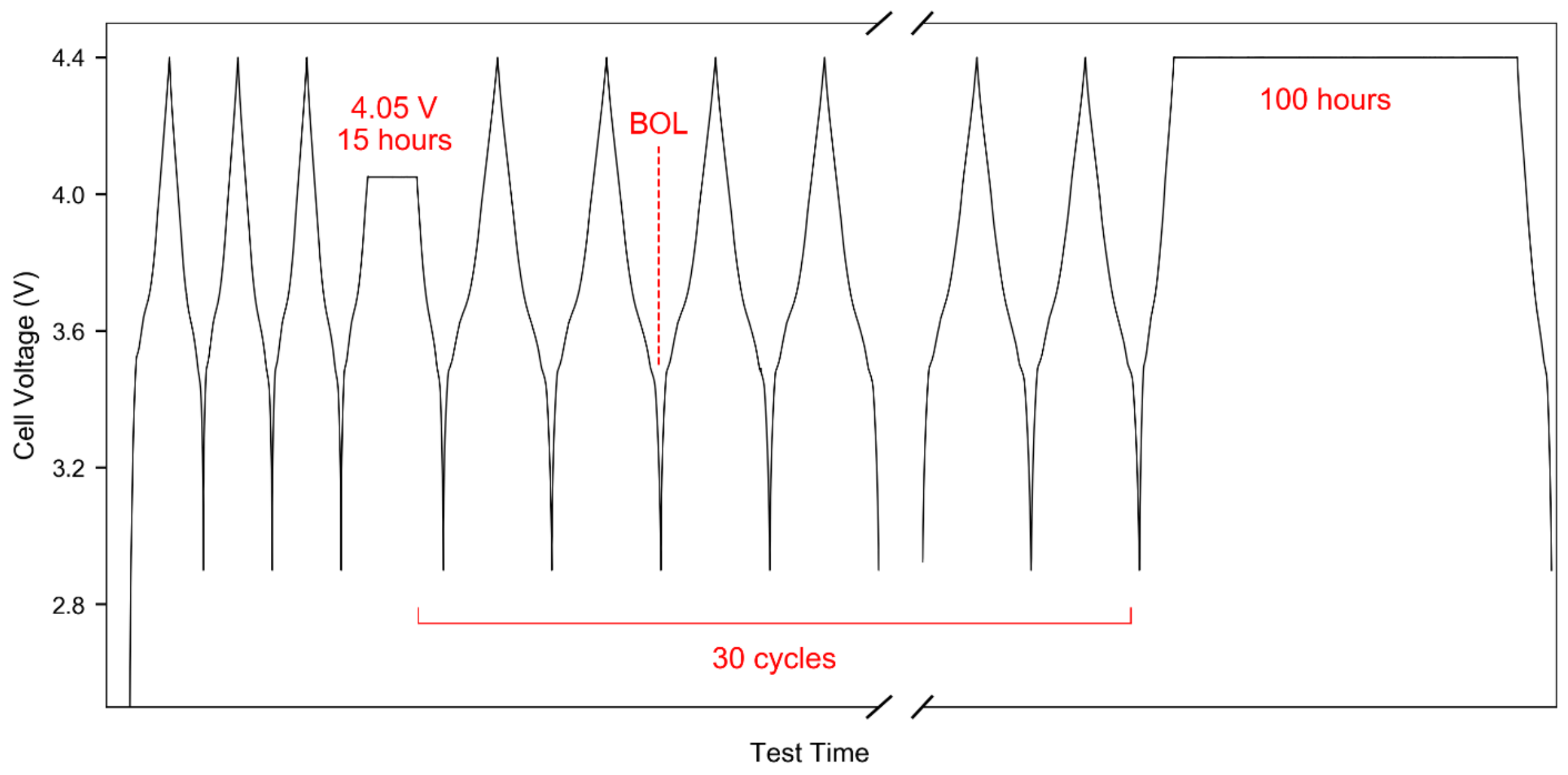

Figure 2. Schematic representation of the testing protocol. The initial six cycles were used for conditioning of the cells and cycle 7 was taken as the beginning of life (BOL). The voltage hold at 4.4 V for 100 hours was used as an additional method for quantifying the rate of parasitic oxidation.

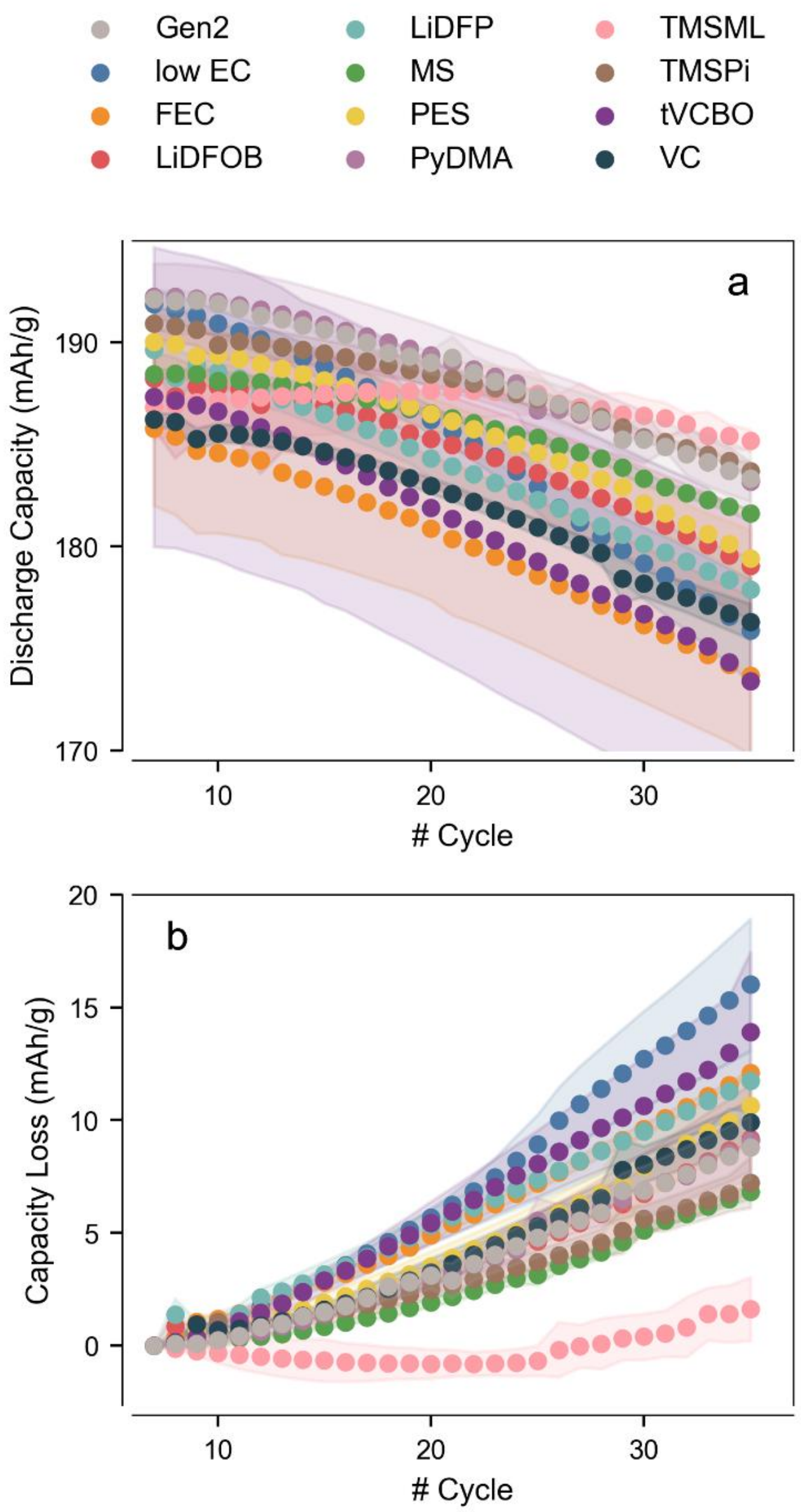

Figure 3. Cycling of cells tested with various electrolyte formulations. a) Discharge capacities. b) Capacity loss. Markers indicate averages from triplicate cells, with shaded areas showing the standard deviation.

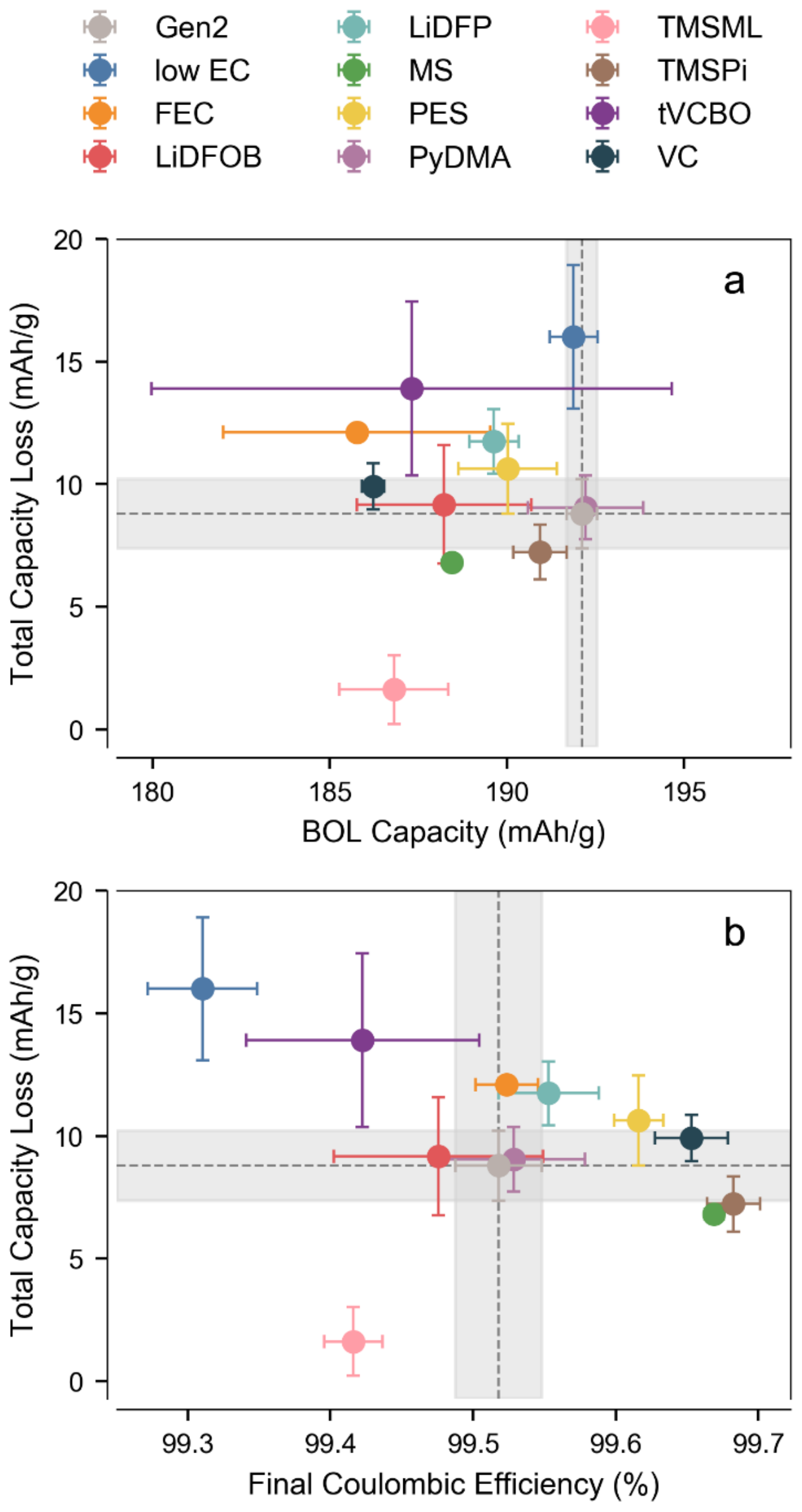

Figure 4. Correlating quantities of interest from the testing data. a) Capacity lost vs. initial capacity. b) Capacity lost vs. the coulombic efficiency at the final cycle. The gray bands demark the standard deviation for the baseline system and provide a visual reference for changes that are statistically significant.

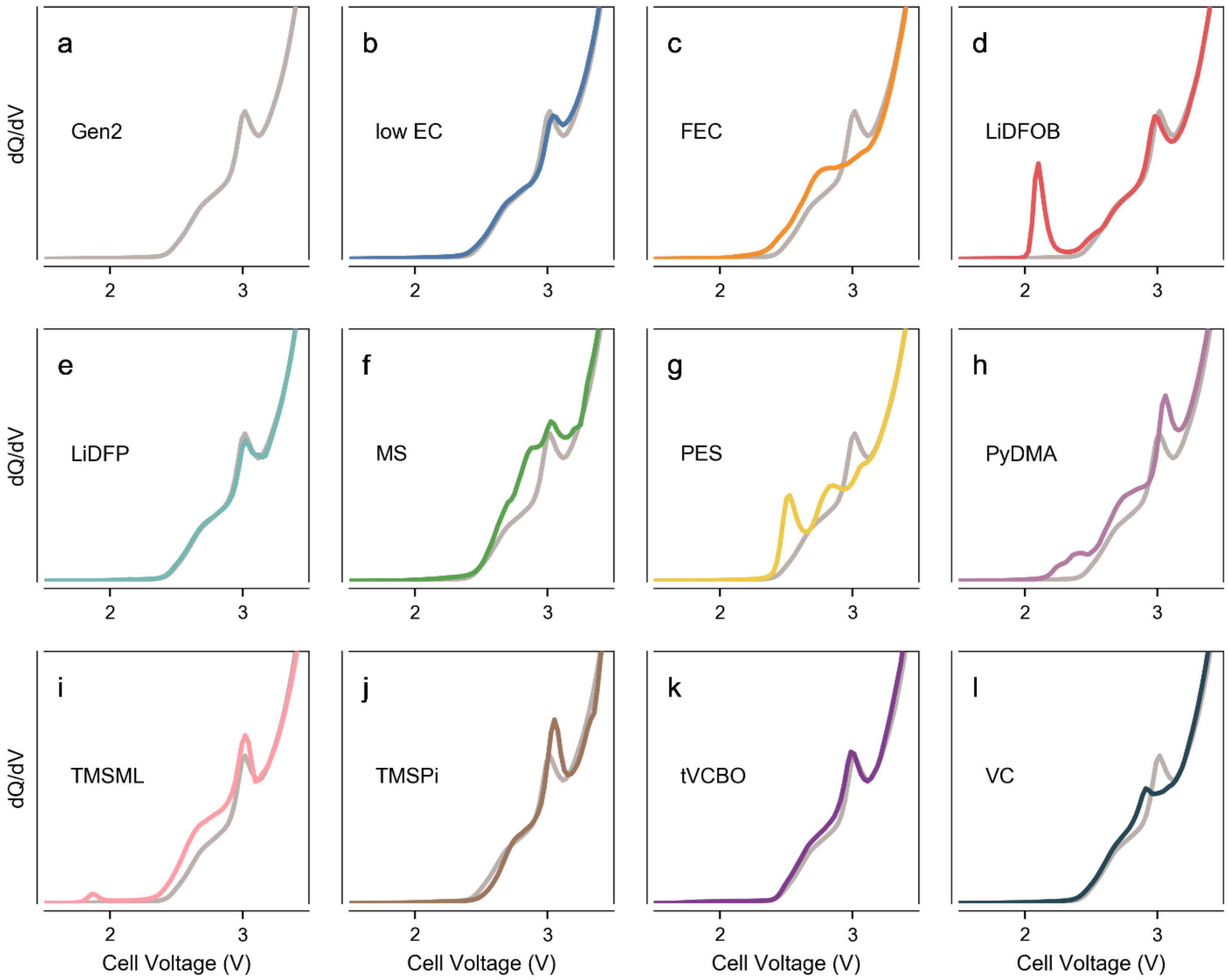

Figure 5. dQ/dV traces during the early stages of formation for all systems. The respective electrolyte formulation is indicated in each panel. Traces for the baseline are replicated in gray for comparison.

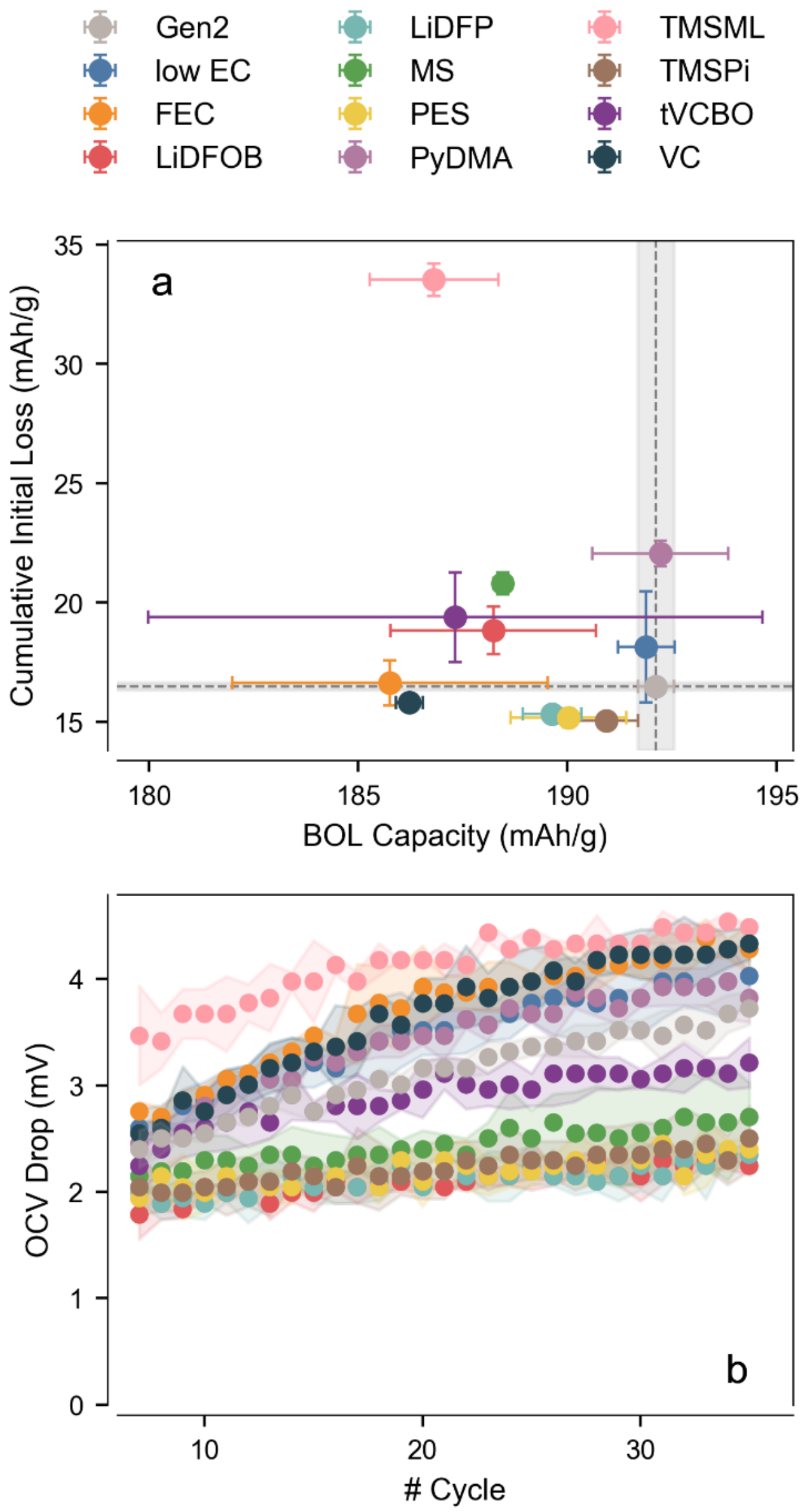

Figure 6. Investigating changes in initial capacity. a) Cumulative loss during the six initial conditioning cycles vs. the initial capacity measured at cycle 7. b) Voltage drop measured during rest at the end of cell charging, assumed to be proportional to impedance.

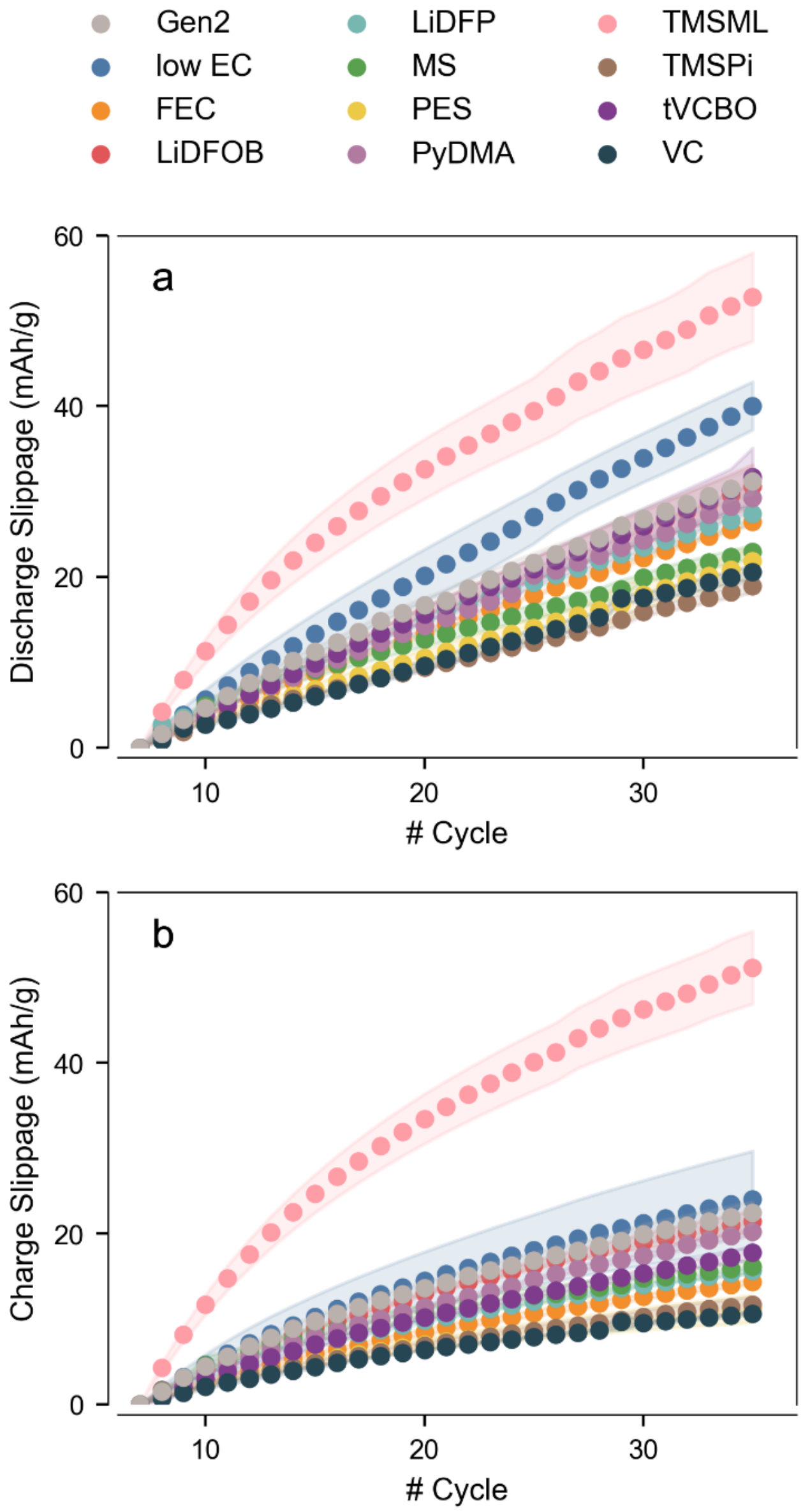

Figure 7. Quantifying side-reactions by tracking the slippage of discharge (a) and charge (b) endpoints. These quantities are proportional to the extent of parasitic reduction and oxidation, respectively. Graphs indicate the cumulative slippage accrued throughout the test.

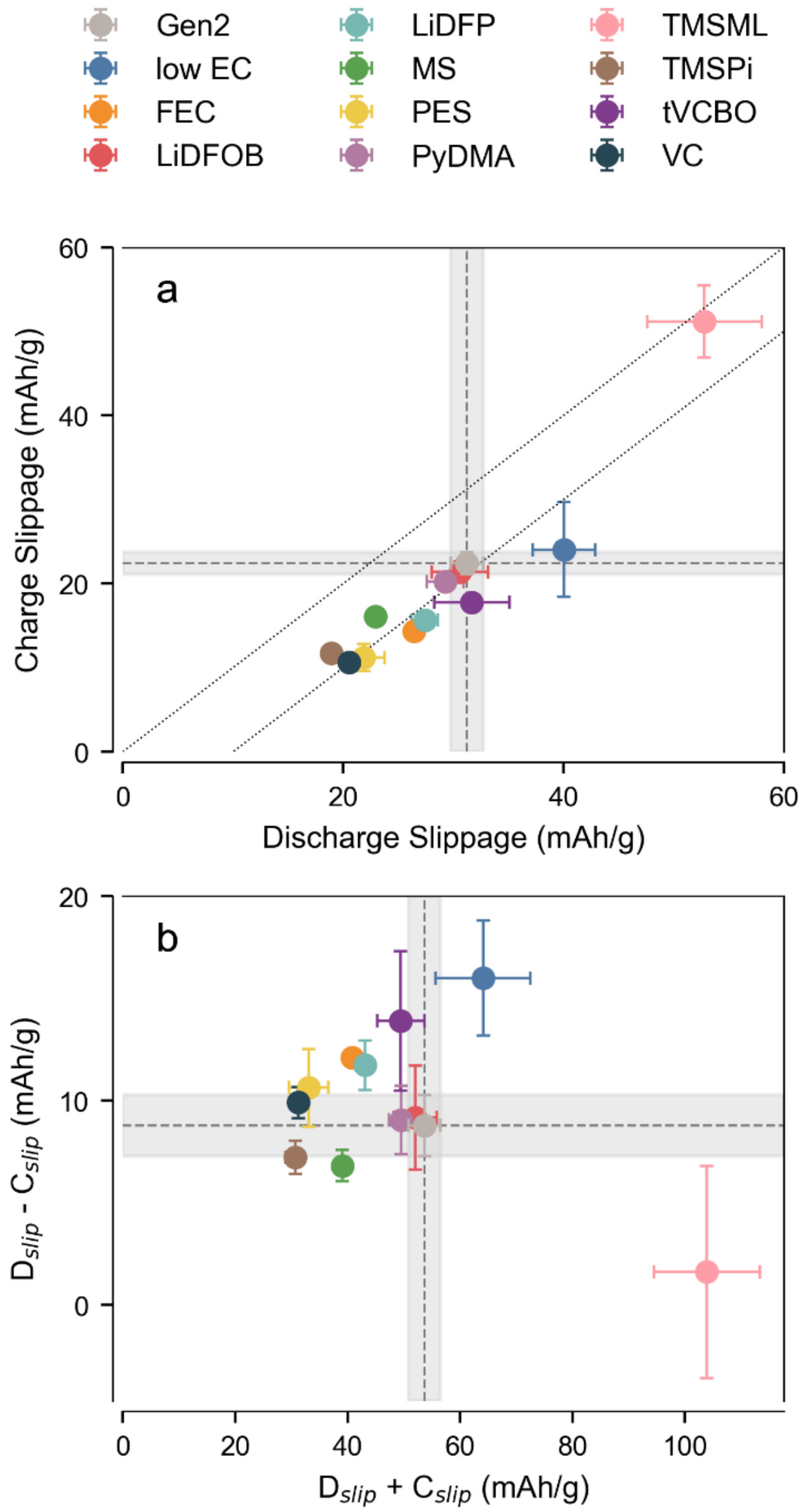

Figure 8. Correlating side-reaction rates. a) Slippage of charge and discharge endpoints accrued during cycling tests. Data points are scattered along a line that is parallel to the main diagonal (also shown for reference). b) Difference between slippages vs. the sum of slippages. The former indicates the balance between capacity lost to and gained from side-reactions, and is equal to the total capacity fade. The latter indicates the total electrolyte degradation resulting from both types of side-reaction. Many additives can lower the reactivity of the cell without improving the capacity retention.

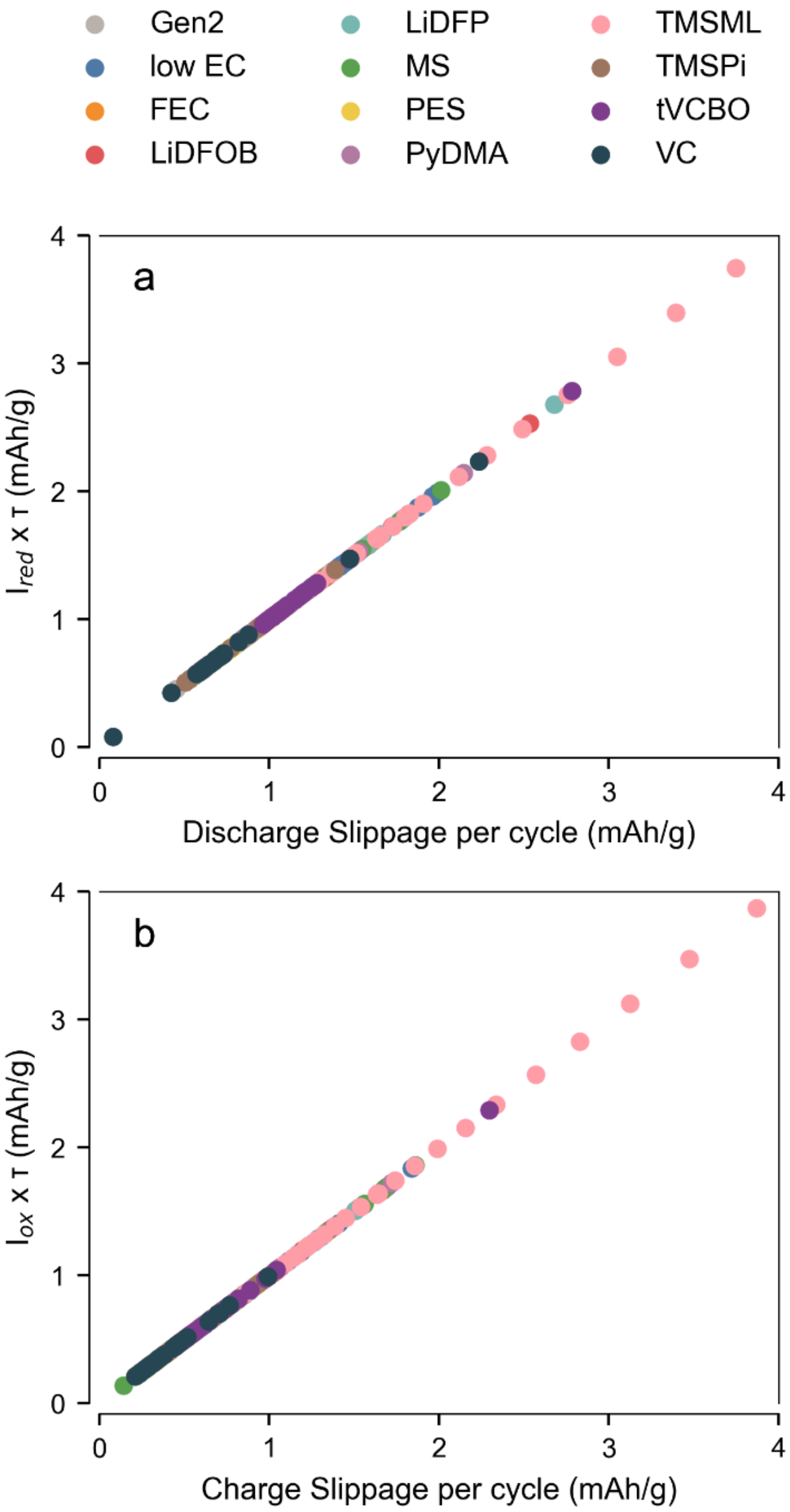

Figure 9. Comparing the parasitic reduction (a) and oxidation (b) quantified through the endpoint slippage and the CE/CR methods. For the latter, side-reaction currents are multiplied by the duration τ of each cycle to yield capacities that can be directly compared with the slippages measured per cycle. Data for all cycles of all systems is presented, showing that both methods provide identical results.

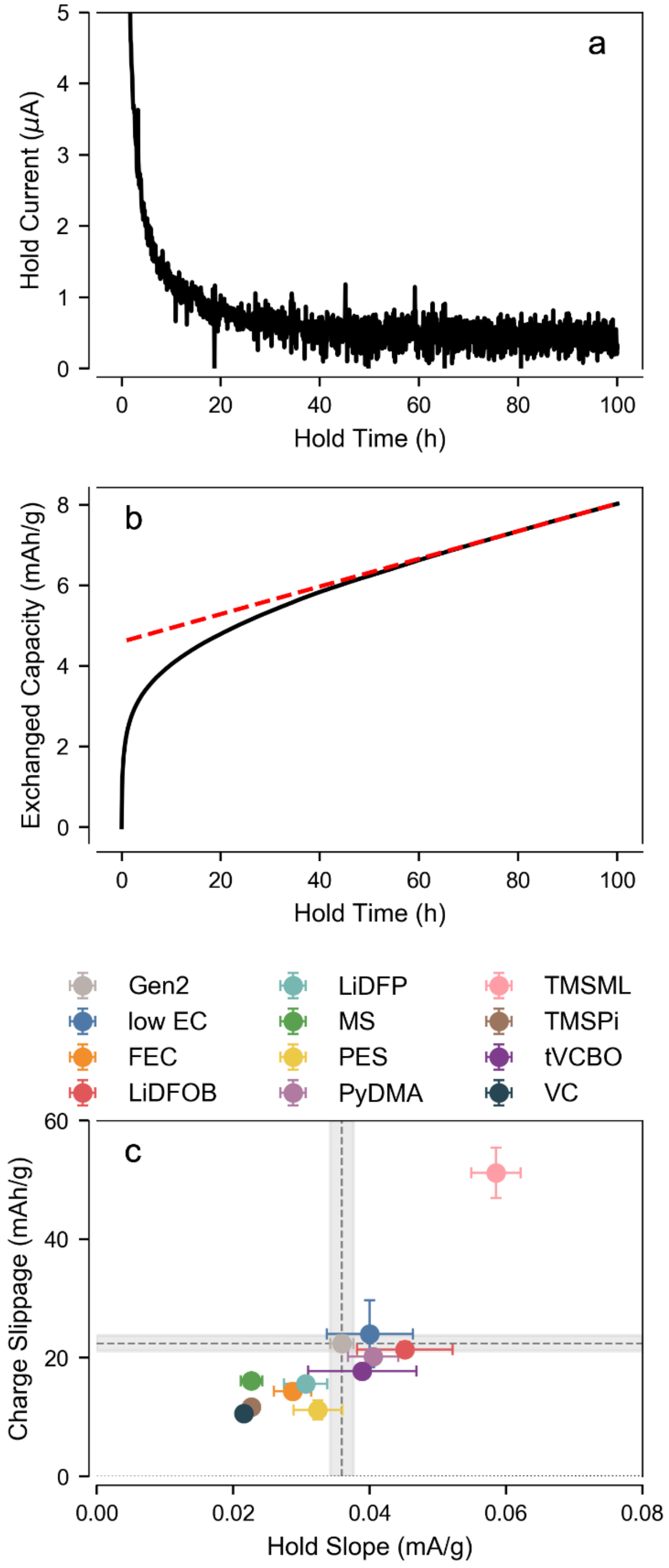

Figure 10. Using voltage holds to quantify parasitic oxidation. a) Example of current measured during the hold. b) Capacity exchanged during the voltage hold. The dashed red trace is a linear fit to the last 30

hours of data that is used for obtaining the parasitic current. c) Comparing voltage hold and charge endpoint slippage. Both methods result in similar general trends.

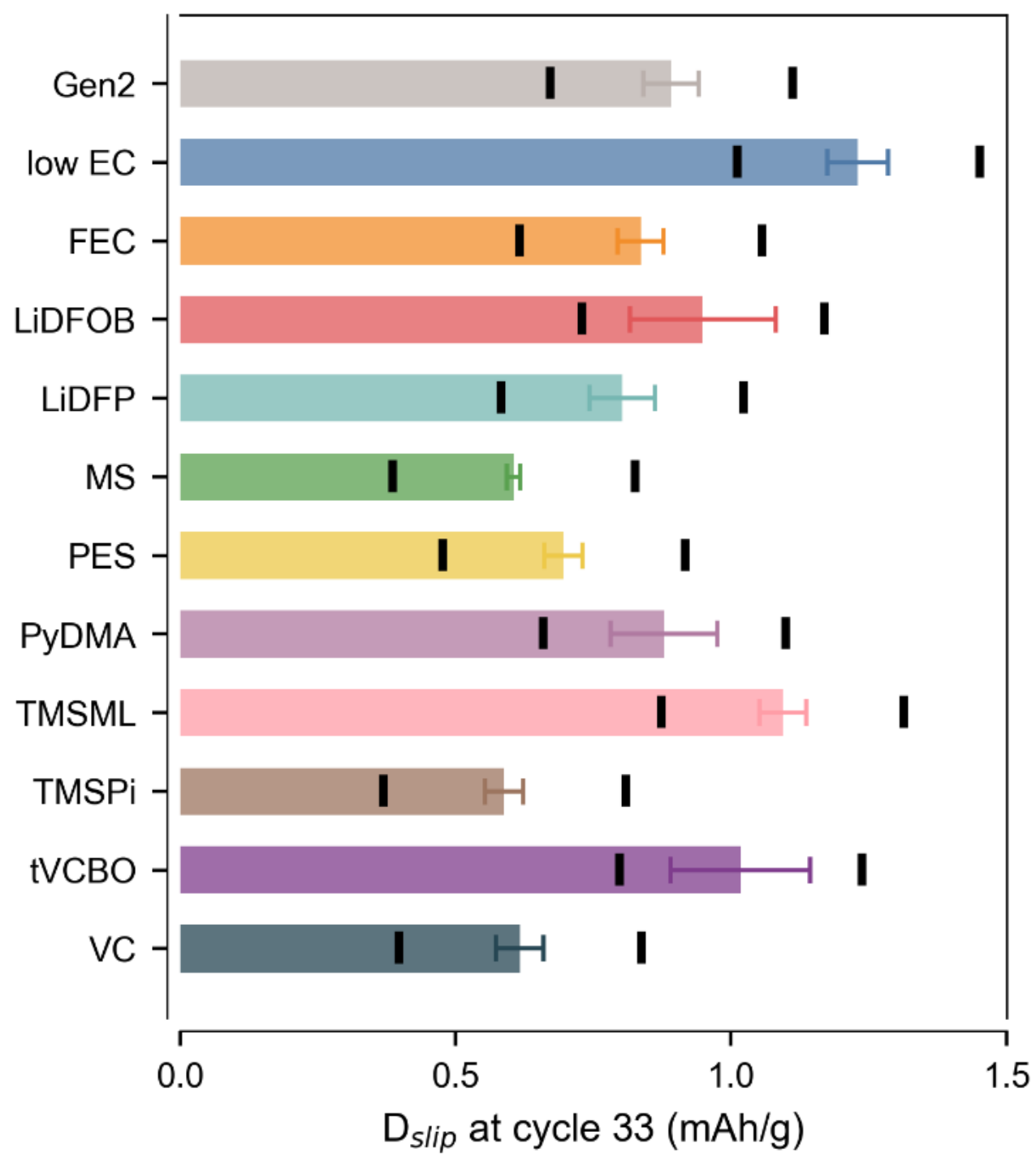

Figure A1. Comparing the estimated hardware error with the dispersion in recorded values. Average discharge endpoint slippages measured at cycle 33 for all electrolyte formulations are depicted in the bar chart. The standard deviation calculated for triplicate cells in each case are shown as error bars with the same color as the corresponding bars in the chart. For reference, the maximum hardware error (3 µAh ≈ 0.22 mAh/g) is shown as bold black traces distributed around the averages for each system. The spread in the data is much tighter than the maximum instrumental uncertainty, suggesting that errors may be systematic.